\documentclass[12pt]{article}
\usepackage{amssymb,amsthm,bm,natbib,bbm}
\usepackage{algorithm}
\usepackage[noend]{algpseudocode}

\usepackage {graphicx}
\usepackage{amsmath,color}
\usepackage{amsfonts}
\usepackage{subfigure}
\usepackage{courier}
\usepackage{verbatim}
\usepackage{setspace}
\usepackage[svgnames]{xcolor}
\usepackage{listings}

\makeatletter
\def\singlespace{\def\baselinestretch{1}\@normalsize}

\newcommand{\vS}{{\bf S}}

\newcommand{\vx}{{\bf x}}
\newcommand{\vy}{{\bf y}}

\newcommand{\vX}{{\bf X}}

\newcommand{\vbeta}{ \bm{\beta}}

\newcommand{\vlam}{\mbox{\boldmath $\lambda$}}

\newcommand{\vxi}{\mbox{\boldmath $\xi$}}

\newcommand{\bay}{\begin{array}}
\newcommand{\eay}{\end{array}}

\newcommand{\vbetah}{\widehat{\vbeta}}

\newcommand{\vbetaw}{\widetilde{\vbeta}}

\newcommand{\ra}{\rightarrow}

\newcommand{\bqa}{\begin{eqnarray*}}
\newcommand{\eqa}{\end{eqnarray*}}
\newcommand{\bqan}{\begin{eqnarray}}
\newcommand{\eqan}{\end{eqnarray}}
\newcommand{\bqt}{\begin{quote}}
\newcommand{\eqt}{\end{quote}}
\newcommand{\bt}{\begin{tabbing}}
\newcommand{\et}{\end{tabbing}}
\newcommand{\bit}{\begin{itemize}}
\newcommand{\eit}{\end{itemize}}
\newcommand{\ben}{\begin{enumerate}}
\newcommand{\een}{\end{enumerate}}
\newcommand{\beq}{\begin{equation}}
\newcommand{\eeq}{\end{equation}}
\newcommand{\bdefi}{\begin{definition}}
\newcommand{\edefi}{\end{definition}}
\newcommand{\bpro}{\begin{proposition}}
\newcommand{\epro}{\end{proposition}}

\newcommand{\bco}{\begin{corollary}}
\newcommand{\eco}{\end{corollary}}
\newcommand{\bdes}{\begin{description}}
\newcommand{\edes}{\end{description}}

\newcommand{\vepsilon}{\mbox{\boldmath $\epsilon$}}

\DeclareMathOperator*{\sumsum}{\sum\sum}

\DeclareMathOperator*{\argmin}{arg\,min}

\def\boxit#1{\vbox{\hrule\hbox{\vrule\kern6pt
          \vbox{\kern6pt#1\kern6pt}\kern6pt\vrule}\hrule}}

\def\bfblue#1{{\color{blue}#1}}

\def\hlblue#1{{\color{blue}#1}}

\mathchardef\ordinarycolon\mathcode`\:
\mathcode`\:=\string"8000
\begingroup \catcode`\:=\active
  \gdef:{\mathrel{\mathop\ordinarycolon}}
\endgroup

\begin{document}

\title{\large
\bf A Survey of Tuning Parameter Selection for High-dimensional Regression}
\author{Yunan Wu and Lan Wang}
\date{}

\maketitle

\begin{singlespace}
\begin{footnotetext}[1]
{Yunan Wu is a Ph.D. candidate and Lan Wang is Professor,  School of Statistics, University of
Minnesota, Minneapolis, MN 55455. Email: wangx346@umn.edu. 
The authors thank the editor’s interest in this topic and an anonymous referee's constructive comments.
They acknowledge the support of VA IIR 16-253 and NSF DMS-1712706. }
\end{footnotetext}
\end{singlespace}

\begin{singlespace}
	\begin{abstract}
		Penalized (or regularized) regression, as represented by Lasso and its variants, has become a standard technique for
		analyzing high-dimensional data when the number of variables substantially exceeds the sample size.
		The performance of penalized regression relies crucially on the choice of the tuning parameter, which
		determines the amount of regularization and hence the sparsity level of the fitted model. The optimal choice
		of tuning parameter depends on both the structure of the design matrix and the unknown random error distribution
		(variance, tail behavior, etc). This article reviews the
		current literature of tuning parameter selection for high-dimensional regression from both theoretical and practical perspectives.
		We discuss various strategies that choose the tuning parameter to achieve prediction accuracy or 
		support recovery. We also review several recently proposed methods for tuning-free high-dimensional
		regression.
	\end{abstract}
\end{singlespace}

\section{Introduction}
High-dimensional data, where the number of covariates/features (e.g., genes) 
may be of the same order or substantially exceed the sample size (e.g., number of patients), 
have become common in many fields due to the advancement in science and technology. Statistical methods for analyzing 
high-dimensional data have been the focus of an enormous amount of research in the past decade or so, 
see the books of \citet{Hastie09}, \citet{buhlmann2011}, \citet{hastie2015statistical} and
\citet{Wainwright19}, among others, for extensive discussions.

In this article, we consider a linear regression model of the form
\begin{align}
\vy = \vX\vbeta_0 + \vepsilon,\label{model}
\end{align} 
where $\vy=(y_1,\ldots,y_n)^T$ is the vector of responses, $\vX = (\vx_1,\ldots,\vx_n)^T$ is an $n\times p$ matrix of covariates, $\vbeta_0=(\beta_{01},\ldots,\beta_{0p})^T$ is the vector of unknown regression coefficients,  
$\vepsilon=(\epsilon_1,\ldots,\epsilon_n)^T$ is a random noise vector with each entry having mean zero and variance $\sigma^2$.  
We are interested in the problem of estimating $\vbeta_0$ when $p\gg n$.
The parameter $\vbeta_0$ is usually not identifiable in high dimension without imposing additional structural assumption, 
as there may 
exist $\vbeta_0'\neq \vbeta_0$ but $\vX\vbeta_0'=\vX\vbeta_0$. 
One intuitive and popular structural assumption underlying a large body of the past work on high-dimensional regression  
is the assumption of strong (or hard) sparsity. Loosely speaking, it means only a relatively small number---usually much less than the sample size $n$---
of the $p$ covariates are active in the regression model.

To overcome the issue of over-fitting, central to high-dimensional data analysis are penalized or regularized
regression techniques represented by Lasso \citep{tibshirani1996, chen2001} and its variants such as Dantzig 
selector \citep{candes2007dantzig},
SCAD \citep{FL:2001}, MCP \citep{Zhang:2010} and Capped $L_1$ \citep{zhang2010analysis}. 
In a nutshell,
a high-dimensional penalized regression estimator solves
\bqan
\label{model1} \min_{\vbeta}\Big\{(2n)^{-1}||\vy-\vX
\vbeta||^2+\sum_{j=1}^pp_{\lambda}(|\beta_j|)\Big\}, \eqan where
$\vbeta=(\beta_1,\ldots,\beta_p)^T$, $||\cdot||$
denotes the $L_2$ vector norm, and $p_{\lambda}(\cdot)$ is a penalty
function which depends on a tuning parameter $\lambda>0$.  Customarily,
the intercept $\beta_0$ is not penalized.

Regardless of the penalty function,
the choice of the tuning parameter $\lambda$ plays a crucial role in the 
performance of the penalized high-dimensional regression estimator. 
The tuning parameter $\lambda$ determines the level of the sparsity of the solution.
Generally speaking, a larger value of $\lambda$ indicates heavier penalty and tends to produce a sparser model.

The paper aims to provide a broad review of the current literature on tuning parameter selection
for high-dimensional penalized regression from both theoretical and practical perspectives.
We discuss different strategies for tuning parameter selection to achieve accurate 
prediction performance or to identify active variables in the model, where the later goal is often 
referred to as support recovery. We also review several recently proposed tuning-free high-dimensional
regression procedures, which circumvent the difficulty of tuning parameter selection.

%An interesting stream of research has investigated how to alleviate the difficulty of selecting $\lambda$ for Lasso. The scaled Lasso of \citet{sun2012scaled} iteratively estimates the regression parameter and $\sigma$; the square-root Lasso (\citet{belloni2014}) eliminates the need to calibrate $\lambda$ for $\sigma$ but does not adjust for the design matrix; TREX (\citet{lederer2015}) automatically adjusts $\lambda$ for both the tail of the error distribution and the design matrix but  the modified loss function is no longer convex;  \citet{sabourin2015} adopts a permutation approach and \cite{chichignoud2016} develops a novel testing procedure to select $\lambda$. These work, however, have not addressed the robustness challenge and may have sub-optimal performance for heavy-tailed errors. For example, the theory of scaled Lasso requires the Gaussian error assumption. Estimating $\sigma$ is particularly challenging for high-dimensional regression with heavy-tailed errors.

\section{Tuning parameter selection for Lasso}
\subsection{Background}
A simple yet successful approach for avoiding over-fitting and enforcing sparsity is to 
regularize the classical least-squares regression with the $L_1$ penalty, corresponding to
adopting $p_{\lambda}(|\beta_j|)=\lambda|\beta_j|$ in (\ref{model1}).
This choice leads to the well known Least Absolute Shrinkage and Selection Operator (Lasso, 
\citet{tibshirani1996}), which simultaneously performs estimation and variable selection.
In the field of signal processing, the Lasso is also known as basis
pursuit \citep{chen2001}.

Formally,
the Lasso estimator $\widehat{\vbeta}^{\scalebox{.7}{\mbox{Lasso}}}(\lambda)$ 
is obtained by minimizing the regularized least squares loss function, that is,
\bqan\label{Lasso}
\widehat{\vbeta}^{\scalebox{.7}{\mbox{Lasso}}}(\lambda)=\argmin_{\vbeta}\Big\{(2n)^{-1}\sum_{i=1}^n(Y_i-\vx_i^T\vbeta)^2+\lambda||\vbeta||_1\Big\},
\eqan
where $\vx_i^T=(x_{i1}, \ldots, x_{ip})$ is the $i$th row of $\vX$,
$||\vbeta||_1$ denotes the $L_1$-norm of $\vbeta$ and $\lambda$ denotes the tuning parameter. 
By varying the value of $\lambda$ and solving the above minimization problem for
each $\lambda$, we obtain a solution path for Lasso.

In the literature, a great deal of work has been devoted to understanding the theoretical 
properties of Lasso, including the theoretical guarantee on the nonasymptotic estimation error bound
$||\widehat{\vbeta}^{\scalebox{.7}{\mbox{Lasso}}}(\lambda)-\vbeta_0||_2$, the prediction
error bound $||\vX(\widehat{\vbeta}^{\scalebox{.7}{\mbox{Lasso}}}(\lambda)-\vbeta_0)||_2$,
and the ability of recovering the support set or the active set of the model 
$\{j: \beta_{0j}\neq 0, j=1, \ldots, p\}$,
see 
\citep{Greenshtein2004, meinshausen2006high, ZY2006, bunea2007, 
van2008, zhang2008sparsity, bickel2009, Candes2009}, among others.
The tremendous success of $L_1$-regularized regression technique is partly due to its computational convenience.
Efficient algorithms such as the exact path-following LARs algorithm \citep{efron2004least} and the fast coordinate descent algorithm
\citep{friedman2007pathwise, wu2008coordinate} have greatly facilitated the
use of Lasso.

\subsection{A theoretical perspective for tuning parameter selection}\label{bear}
Motivated by the Karush-Kuhn-Tucker condition for convex optimization \citep{boyd2004},  
\citet{bickel2009} proposed a general principal for selecting $\lambda$
for Lasso. More specifically, it is suggested that
$\lambda$ should be chosen such that 
\bqan\label{toptimal}
P\Big\{||n^{-1}\vX^T\vepsilon||_{\infty}\leq \lambda\Big\}\geq 1-\alpha,
\eqan
for some small $\alpha>0$, 
where  $||\cdot||_{\infty}$ denotes the infinity (or supremum) norm.

Consider the important example where the random errors $\epsilon_i$, $i=1, \ldots, n$, are independent $N(0,\sigma^2)$
random variables and the design matrix is normalized such that each column has $L_2$-norm 
equal to $\sqrt{n}$. One can show that an upper bound of $\lambda$ satisfying  (\ref{toptimal}) is given by 
$\tau \sigma\sqrt{\log p/n}$ for some positive constant $\tau$. To see this, we observe that by the property
of the tail probability of Gaussian distribution and the union bound, 
\bqa\label{toptimal}
P\Big\{||n^{-1}\vX^T\vepsilon||_{\infty}\leq \tau \sigma\sqrt{\log p/n}\Big\}\geq 1-2\exp\big(-(\tau^2-2)\log p/2\big),
\eqa
for some $\tau>\sqrt{2}$. Similar probability bound holds if the random errors
have sub-Gaussian distributions (e.g., Section 4.2 of \citep{Negahban2012}).

Most of the existing theoretical properties of Lasso were derived 
while fixing $\lambda$ at 
an oracle value satisfying (\ref{toptimal}) or within a range of oracle values whose bounds satisfying similar constraints.
For example, the near-oracle error bound of Lasso given in \citet{bickel2009}
was derived assuming $\lambda=\tau \sigma\sqrt{\log p/n}$ for some $\tau>2\sqrt{2}$ 
when $\vX$ satisfies a restricted eigenvalue condition.
See \citet{buhlmann2011} for further discussions on the restricted eigenvalue condition
and other similar conditions on $\vX$ to guarantee that the design matrix is well behaved
in high dimension.

%In fact, careful analysis of the theory of Lasso suggests that $\lambda$ is an important
%factor appearing in the estimation error bound of Lasso. For optimality, it is desirable to choose
%the smallest $\lambda$ such that (\ref{toptimal}) holds.

The theory of Lasso suggests that $\lambda$ is an important
factor appearing in its estimation error bound. To achieve optimal estimation error bound, it is desirable to choose
the smallest $\lambda$ such that (\ref{toptimal}) holds. This choice, however, depends on both the unknown random error distribution
and the structure of the design matrix $\vX$. As discussed above,
a reasonable upper bound for such a theoretical choice of $\lambda$ requires the knowledge of
$\sigma$, the standard deviation of the random error.
Estimation of $\sigma$ in high dimension is itself a difficult problem.
As a result, it is often infeasible to apply the theoretical choice of $\lambda$ 
in real data problems.

\subsection{Tuning parameter selection via cross-validation}

In practice, a popular approach to selecting the tuning parameter $\lambda$ for Lasso
is a data-driven scheme called cross-validation, which aims for optimal prediction accuracy.
Its basic idea is to randomly split the data into a training data set and a testing (or validation) data set such that
one may evaluate the prediction error on the testing data while fitting the model using the training data set.
%Its basic idea is the following: to divide the whole dataset into two sets, which are the training dataset $(\vy_T,\vX_T)$, and the validation %dataset $(\vy_V,\vX_V)$. Then with any given tuning parameter $\lambda$, we estimate $\vbetah(\lambda)$ from (\ref{model2}) using the training %dataset. Then with this Lasso estimator, we can compute the mean squared of prediction errors given the validation dataset. 
There exist several different versions of cross-validation, such as leave-$k$-out cross-validation, repeated random sub-sampling validation (also known as Monte Carlo cross-validation), and $K$-fold cross-validation. Among these options, $K$-fold cross-validation is most
widely applied in real-data analysis.

The steps displayed in Algorithm~\ref{algo:kfoldcv} illustrate the implementation of $K$-fold cross-validation for Lasso. 
The same idea broadly applies to more general problems such as penalized likelihood estimation with different penalty functions.
First, the data is randomly partitioned into $K$ roughly equal-sized subsets (or folds), where typical choice of $K$ is 5 or 10. 
Given a value of $\lambda$,
one of the $K$ folds is retained as the validation data set to evaluate the prediction error, and the remaining data 
are used as the training data set to obtain $\widehat{\vbeta}^{\scalebox{.7}{\mbox{Lasso}}}(\lambda)$.
This cross-validation process is then repeated, with each of the $K$ folds being used as the validation data set exactly once. For example, in carrying out a 5-fold cross-validation for Lasso, we randomly split the data into five roughly equal-sized parts $\mathcal{V}_1,\cdots,\mathcal{V}_5$. Given a tuning parameter $\lambda$, we first train the model and estimate $\widehat{\vbeta}^{\scalebox{.7}{\mbox{Lasso}}}(\lambda)$ on $\{\mathcal{V}_2,\cdots,\mathcal{V}_5\}$ and then compute the 
total prediction error on $\mathcal{V}_1$. Repeat this process by training on $\{\mathcal{V}_1, \mathcal{V}_3,\mathcal{V}_4,\mathcal{V}_5\}$ and validating on $\mathcal{V}_2$, and so on. The cross-validation error $\mbox{CV}(\lambda)$ is obtained as the average of the prediction errors over the $K$
validation data sets from this iterative process.    

\begin{algorithm}[!h] 
	\caption{K-fold cross-validation for Lasso} \label{algo:kfoldcv} 
	\begin{algorithmic}[1]
		\State Randomly divide the data of sample size $n$ into $K$ folds, $\mathcal{V}_1, . . . \mathcal{V}_K$, of roughly equal sizes.
		\State Set $\mbox{Err}(\lambda)$ = 0.
		\For {$k=1,\cdots,K $} 
		\State  Training dataset $(\vy_T,\vX_T)=\{(y_i,\vx_i): i\notin \mathcal{V}_k\} $.
		\State  Validation dataset $(\vy_V,\vX_V)=\{(y_i,\vx_i): i\in \mathcal{V}_k\}$.
		\State $\vbetah^{\scalebox{.7}{\mbox{Lasso}}}(\lambda)\leftarrow \argmin\limits_{\vbeta} \big\{(2|\mathcal{V}_k|)^{-1}||\vy_T-\vX_T
		\vbeta||^2+\lambda||\vbeta||_1\big\}$. 
		\State $\mbox{Err}(\lambda) \leftarrow\ \mbox{Err}(\lambda) + ||\vy_V-\vX_V \vbetah^{\scalebox{.7}{\mbox{Lasso}}}(\lambda)||^2$.	
		\EndFor
		\Return  $\mbox{CV}(\lambda)=n^{-1}\mbox{Err}(\lambda)$.	
	\end{algorithmic}
\end{algorithm}

% \begin{figure}
% 	\centering
% 	\includegraphics[width=0.9\linewidth,height=0.4\textheight]{KfoldCV}
% 	\caption{5-fold cross-validation}
% 	\label{fig:kfoldcv}
% \end{figure}
Given a set $\Lambda$ of candidate tuning parameter values, say a grid $\{\lambda_1, \ldots, \lambda_m\}$, one would compute $\mbox{CV}(\lambda)$ according to Algorithm~\ref{algo:kfoldcv} for each $\lambda\in\Lambda$. This yields the cross-validation error curve $\{\mbox{CV}(\lambda): \lambda\in\Lambda\}$.
To select the optimal $\lambda$ for Lasso, two useful strategies are usually recommended.
A simple and intuitive approach is to select the $\lambda$ that minimizes the cross-validation error, i.e.,
\bqan\label{minl}
\widehat{\lambda} = \argmin\limits_{\lambda} \mbox{CV}(\lambda).\label{cv_crit}
\eqan
An alternative strategy is based on the so-called “one-standard-error rule”, which
chooses the most parsimonious model (here corresponding to larger $\lambda$ and more regulation)  
such that its cross-validation error is within one standard-error of 
$\mbox{CV}(\widehat{\lambda} )$.
This is feasible as the $K$-fold cross-validation allows 
one to estimate the standard error of the cross-validation error.
The second strategy acknowledges
that the cross-validation error curve is estimated with error and is motivated by the principle of parsimony
(e.g., Section 2.3 of \citet{hastie2015statistical}).

%Another option  for choosing the value of the tuning parameter is the \textbf{one standard error rule}. In this way, %we also find the usual minimizer $\widehat{\lambda}$ as above, and then move $ \widehat{\lambda}$ in the direction of %increasing regularization (this may be increasing or decreasing $\lambda$, depending on the parametrization) as much %as we can, such that the cross-validation error curve is still within one standard error of $Err(\widehat{\lambda} )$. %In other words, we select $\lambda$ that achieves the sparsest model and $Err(\lambda)$ is  equal (up to one standard %error) to $Err(\widehat{\lambda})$.

%``All else equal (up to one standard error), go for the simpler (more regularized) model''
Several R functions are available to implement $K$-fold cross-validation for Lasso, such as the ``cv.glmnet''
function in the R package \texttt{glmnet} \citep{glmnet} and the``cv.lars''function in the R package \texttt{lars} \citep{lars}. 
Below are the sample R codes for performing the $5$-fold cross-validation for Lasso using the ``cv.glmnet'' function.
\begin{lstlisting}
library(glmnet)
data(SparseExample) 
cvob1=cv.glmnet(x, y, nfolds=5)
plot(cvob1)
\end{lstlisting}
\begin{figure}
	\centering
	\includegraphics[width=0.7\linewidth]{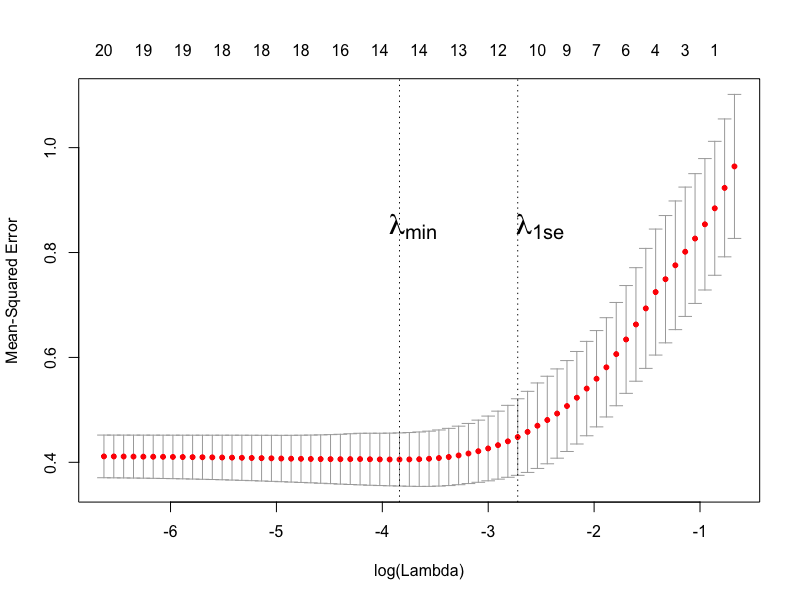}
	\caption{Cross-validation for Lasso}
	\label{fig:cvglmnet}
\end{figure}
The plot produced by the above commands is given in 
Figure~\ref{fig:cvglmnet}, which depicts the cross-validation error curve 
(based on the mean-squared prediction error in this example) as well as the one-standard-error
band. In the plot, $\lambda_{\min}$ is the tuning parameter
obtained by (\ref{minl}), i.e., the value of the tuning parameter
 that minimizes the cross-validation prediction error. And $\lambda_{1se}$ denotes the tuning parameter selected via the one-standard-error rule. The numbers at the top of the plot correspond to the numbers of non-zero coefficients
 (or model sizes) for models fitted with different $\lambda$ values. For this data example, 
 the prediction errors based on $\lambda_{\min}$ and $\lambda_{1se}$ are close, while the model based on 
 $\lambda_{1se}$ is notably sparser than the one based on $\lambda_{\min}$.

In the existing work on the theory of Lasso, the tuning parameter is usually considered to be deterministic,
or fixed at a pre-specified theoretical value. Despite the promising
empirical performance of cross-validation, much less is known about
the theoretical properties of the cross-validated
Lasso, where the tuning parameter is selected in a data-driven manner.
Some important progress has been made recently in understanding the 
properties of cross-validated Lasso. \cite{Homrighausen2013},
\cite{Chatterjee2015} and \cite{Homrighausen2017}
investigated the risk consistency of cross-validated Lasso under different regularity 
conditions.
 \cite{Chetverikov2016} derived a nonasymptotic error bound for 
  cross-validated Lasso and showed that it can achieve the optimal estimation rate 
  up to a factor of order $\sqrt{\log(pn)}$.

\subsection{Scaled Lasso: adapting to unknown noise level}
As discussed earlier, the optimal choice of $\lambda$ for Lasso requires the knowledge
of the noise level $\sigma$, which is usually unknown in real data analysis.
Motivated by \citet{stadler2010} and the discussions on this paper by
\citet{antoniadis2010comments} and \citet{sun2010comments}, 
\citet{sun2012scaled} thoroughly investigated the performance of an iterative algorithm
named scaled Lasso, which jointly estimates the 
regression coefficients $\vbeta_0$ and the noise level $\sigma$ in a sparse
linear regression model.

Denote the loss function for Lasso regression by
\begin{align}
L_\lambda(\vbeta) = (2n)^{-1}||\vy-\vX \vbeta||^2+\lambda||\vbeta||_1. \label{scaled_lasso_loss}
\end{align}
The iterative algorithm for scaled Lasso is described in Algorithm~\ref{algo:scaledLasso}, 
where $\vbeta^0$ and $\lambda_0$ are initial values independent of $\vbeta_0$ and $\sigma$. 
In this algorithm, the tuning parameter is rescaled iteratively.
In the output of the algorithm, $\vbetah(\vX,\vy)$ is referred to as the scaled Lasso estimator.
\begin{algorithm}[!h] 
	\caption{Scaled Lasso algorithm (Sun and Zhang, 2012)} \label{algo:scaledLasso} 
	\begin{algorithmic}[1]
		\State Input $(\vX,\vy)$, $\vbeta^0$, and $\lambda_0$.
		\State $\vbeta\leftarrow \vbeta^{0}$.
		\While {$L_\lambda(\vbeta)\leq L_\lambda(\vbeta^{0})$} 
		\State $\vbeta^{0}\leftarrow \vbeta$
		\State $\widehat{\sigma} \leftarrow n^{-1/2}||\vy-\vX\vbeta^{0}||$.
		\State $\lambda \leftarrow \widehat{\sigma}\lambda_0$.
		\State $\vbeta  \leftarrow \argmin\limits_{\vbeta}  L_\lambda(\vbeta) $.
		\EndWhile
		\Return  $\widehat{\sigma}(\vX,\vy)\leftarrow \widehat{\sigma} $ and $\vbetah(\vX,\vy)\leftarrow \vbeta^0$.	
	\end{algorithmic}
\end{algorithm}

\citet{sun2012scaled} showed that the outputs of Algorithm~\ref{algo:scaledLasso} 
converge to the solutions of a joint minimization problem, specifically,
\begin{align}
(\vbetah,\widehat{\sigma}) %= \argmin\limits_{\vbeta,\sigma}L_{\lambda_0}(\vbeta,\sigma) 
= \argmin\limits_{\vbeta,\sigma} \Big\{(2n\sigma)^{-1}||\vy-\vX \vbeta||^2 +\sigma/2+ \lambda_0||\vbeta||_1\Big\}. \label{scaled_lasso}
\end{align}
This is equivalent to jointly minimizing
Huber’s concomitant loss function with the $L_1$ penalty \citep{Owen2007, antoniadis2010comments}. 
This loss function possesses the nice property of being jointly convex in $(\vbeta, \sigma)$.
It can also be shown that the solutions are scale-equivariant in $\vy$, i.e., 
$\vbetah(\vX,c\vy) = c\vbetah(\vX,\vy)$ and $\widehat{\sigma}(\vX,c\vy) = |c|\widehat{\sigma}(\vX,\vy)$ for any constant $c$. This property is practically important in data analysis. 
Under the Gaussian assumption and other mild regularity conditions, \citet{sun2012scaled} derived oracle inequalities
for prediction and joint estimation of $\sigma$ and $\vbeta_0$ for the scaled
lasso, which in particular imply the consistency and asymptotic normality of $\widehat{\sigma}(\vX,\vy)$ as an
estimator for $\sigma$.

The function ``scalreg'' in the R package \texttt{scalreg} implements Algorithm~\ref{algo:scaledLasso} for the scaled Lasso. 
The sample codes below provide an example on
how to analyze the ``sp500'' data in that package with scaled Lasso.

\begin{lstlisting}
library(scalreg)
data(sp500) 
attach(sp500)
x = sp500.percent[,3: (dim(sp500.percent)[2])]
y = sp500.percent[,1]
scaleob <- scalreg(x, y)
\end{lstlisting}

\section{Alternative $L_1$-penalty based methods: from tuning selection to tuning free}
This section provides a brief review of several recently proposed
$L_1$-penalty based tuning-free procedures for high-dimensional sparse linear regression. 
These procedures
tackle the challenge of tuning parameter selection for Lasso from different angles.
As suggested by (\ref{toptimal}), the theoretically optimal tuning parameter for Lasso 
depends on both the design matrix $\vX$ and the unknown error distribution
(standard deviation $\sigma$, tail behavior, etc).
The three procedures we review here (square-root Lasso, TREX
and Rank Lasso) aim to automatically adapt to one or more aspects of these factors.

\subsection{Scale-free square-root Lasso}
Square-root Lasso is a variant of Lasso proposed by \citet{belloni2011} which enjoys the advantage to
avoid calibrating the tuning parameter with respect to the noise level $\sigma$.  
Square-root Lasso replaces least squares loss (or $L_2$ loss) function in Lasso with 
its positive square root. 
Assuming the $\epsilon_i$ are independently distributed with mean zero and variance $\sigma^2$,
the square-root Lasso estimator is defined as
\begin{align}
\label{sqrtLasso}
\vbetah_{\sqrt{\scalebox{.7}{\mbox{Lasso}}}}(\lambda) =\argmin\limits_{\vbeta}\Big\{n^{-1/2}||\vy-\vX\vbeta|| + \lambda||\vbeta||_1\Big\}.
\end{align}

Let $L_{\scalebox{.7}{\mbox{SR}}}(\vbeta)=n^{-1/2}||\vy-\vX\vbeta||$ denote the loss function
of square-root Lasso and let $S_{\scalebox{.7}{\mbox{SR}}}$ denote its 
subgradient evaluated at $\vbeta=\vbeta_0$.
The general principal of tuning parameter selection 
(e.g., \citet{bickel2009}) suggests to choose $\lambda$ such that
$
P(\lambda>c||S_{\scalebox{.7}{\mbox{SR}}}||_{\infty})\geq 1-\alpha_0,
$
for some constant $c>1$ and a given small $\alpha_0>0$. 
An important observation that underlies the advantage of square-root Lasso is that
\bqa
S_{\scalebox{.7}{\mbox{SR}}}=\frac{n^{-1}\sum_{i=1}^n\vx_i\epsilon_i}{(n^{-1}\sum_{i=1}^n\epsilon_i^2)^{1/2}}
\eqa
does not depend on $\sigma$.

Computationally, the square-root lasso can be formulated as a solution to a convex conic programming problem. The function ``slim'' in the R package \texttt{flare} \citep{flare} implements a family of Lasso variants for high-dimensional regression, including the square-root Lasso. The sample codes below demonstrate how to 
implement the square-root Lasso using this function to
analyze the ``sp500'' data in the \texttt{scalreg} package. 
The arguments \texttt{\hlblue{method="lq", q = 2}} 
yield square-root Lasso, which are also the default options in the ``slim'' function.

\begin{lstlisting}
library(flare)
data(sp500) 
attach(sp500)
x = sp500.percent[,3: (dim(sp500.percent)[2])]
y = sp500.percent[,1]
sqrtob <- slim(x, y, method="lq", q = 2)
\end{lstlisting}

\citet{belloni2011} recommended the choice 
$\lambda =c n^{-1/2}\Phi^{-1}\Big(1-\frac{\alpha}{2p}\Big)$,
for some constant $c>1$ and $\alpha>0$. Note that this choice of $\lambda$ does not depend on $\sigma$, and it is valid asymptotically without requiring the random errors to be Gaussian.
Under general regularity conditions, \citet{belloni2011} showed that
there exists some positive constant $C_n$ such that
$$P\Big(||\vbetah_{\sqrt{\scalebox{.7}{\mbox{Lasso}}}}(\lambda) -\vbeta_0||\leq  C_n\sigma\big\{n^{-1}s\log(2p/\alpha)\big\}^{1/2}\Big) \geq  1-\alpha,$$
where $s=||\vbeta_0||_0$ is the sparsity size of the true model. 
Hence, square root Lasso achieves the near-oracle
rate of Lasso even when $\sigma$ is unknown. 

The square-root Lasso and Lasso are equivalent families of estimators. 
There exists a one-to-one mapping between the tuning
parameter paths of square-root Lasso and Lasso \citep{tian2018selective}.
It is also worth pointing out that
the square-root Lasso is related to but should not be
confused with the scaled Lasso \citep{sun2012scaled}.
The current literature contain some confusion (particularly in the use of names) 
about these two methods.
The connection and distinction
between them are nicely discussed in Section 3.7 of 
\cite{van2016estimation}.

%It establishes the high probability bound for $L_2$ error of square-root Lasso estimator. If the  %noise vector $\vepsilon$ is non-normal, \citet{belloni2011} also proposed three options for %$\lambda$, which are the exact, semi-exact and asymptotic options. With any of these options for %$\lambda$ selection, similar error bounds can be derived, with mild restrictions on the design %matrix $\vX$ and the tail behavior of $\vepsilon$.

%The theoretical results imply that the optimal $\lambda$ for the square-root Lasso estimator is of %the order $$\lambda_{opt}\sim \frac{||\vX^T\vepsilon||_\infty}{n},$$
%which is pivotal with respect to the noise level $\sigma$. It requires neither the known $\sigma$, %nor the pre-estimator of $\sigma$, and matches the near-oracle performance of the Lasso with known %$\sigma$. It also suggests that square-root Lasso estimators are scale equivariant in $\vy$, like %the scaled Lasso estimator. 

\subsection{TREX} 
The scaled Lasso and square-root Lasso both address the need to calibrate $\lambda$ for $\sigma$. However, the tail behavior of the noise vector, as well as the structure of the design matrix, could also have significant effects on the optimal selection of $\lambda$. 
To alleviate these additional difficulties, \citet{lederer2015} proposed a new approach for high-dimensional variable selection.
The authors named the new approach TREX to emphasize that it aims at
Tuning-free Regression that adapts to the Entire noise and the
design matrix $\vX$.
Indeed, the most attractive property of TREX is that it automatically adjusts $\lambda$ for
the unknown noise standard deviation 
$\sigma$, the tail of the error distribution and the design matrix.  

In contrast to the square-root Lasso, the TREX estimator modifies the Lasso loss function in a different way.
The TREX estimator is defined as
\begin{align}
\label{TREX} 
\vbetah_{\scalebox{.7}{\mbox{TREX}}} =\argmin\limits_{\vbeta}\Big\{L_{\mbox{TREX}}(\vbeta) + ||\vbeta||_1\Big\}, 
\end{align}
where 
\bqa
L_{\scalebox{.7}{\mbox{TREX}}}(\vbeta) = \frac{2||\vy-\vX \vbeta||^2}{||\vX^T(\vy-\vX\vbeta)||_\infty}.
\eqa
TREX does not require a tuning parameter.
In this sense, it is a completely tuning-free procedure.
\citet{lederer2015} proved that the TREX estimator 
 is close to a Lasso solution with tuning parameter of the same order as the theoretically optimal $\lambda$. They presented examples where TREX 
 has promising performance comparing with Lasso.
 
The modified loss function for the TREX estimator, however, is no longer convex. 
\citet{bien2016non} showed the remarkable result that
despite the non-convexity, there exists a polynomial-time algorithm that is guaranteed to find the global minimum of the TREX problem. 
\citet{bien2018prediction} recently established a prediction
error bound for TREX, which deepens the understanding of the theoretical properties of TREX.

\subsection{Rank Lasso: a tuning free and efficient procedure} 
Recently, \citet{Wang2018} proposed an alternative approach to overcoming the challenges of tuning
parameter selection for Lasso. The new method, named Rank Lasso, has an optimal tuning parameter 
that can be easily simulated
and automatically adapts to both the unknown
random error distribution and the structure of the design matrix.
Moreover, it enjoys several other appealing properties: 
it is a solution to a convex optimization problem and can be
conveniently computed via linear programming; it has similar performance as Lasso does
when the random errors are normally distributed and is
robust with substantial efficiency gain for heavy-tailed random errors;
it leads to a scale-equivariant estimator
which permits coherent interpretation when the response variable undergoes a
scale transformation.

Specifically, the new estimator is defined as
\bqan\label{dragon}
\vbetah_{\scalebox{.7}{\mbox{rank}}}(\lambda) = \argmin\limits_{\vbeta}\Big\{Q_n(\vbeta) +\lambda||\vbeta||_1\Big\}, \label{rank_est}
\eqan
where the loss function 
\begin{align}
Q_n(\vbeta) = [n(n-1)]^{-1} \sumsum_{i\neq j}\big|(y_i-\vx_i^T\vbeta)-(y_j-\vx_j^T\vbeta)\big|. \label{rank_loss}
\end{align}
The loss function $Q_n(\vbeta)$ is related to 
Jaeckel's dispersion function with Wilcoxon scores \citep{jaeckel1972}
in the classical nonparametric statistics literature.
For this reason, the estimator in (\ref{dragon}) is referred to
as the rank Lasso estimator. 
In the classical low-dimensional setting, 
regression with Wilcoxon loss function was investigated by \citet{wang2009}
and \citet{wang2009local}.

To appreciate its tuning free property, we observe that the gradient function of 
$Q_n(\vbeta)$ evaluated at $\vbeta_0$ is 
\bqa
\vS_n := \frac{\partial Q_n(\vbeta) }{\partial \vbeta}\Big|_{\vbeta=\vbeta_0}=-2[n(n-1)]^{-1}\vX^T\vxi,
\eqa
where $\vxi = (\xi_1,\cdots,\xi_n)^T$ with $\xi_i = 2r_i - (n+1)$ and 
$r_i = \mbox{rank}(\epsilon_i)$ among $\epsilon_1,\cdots,\epsilon_n$.
Note that the random vector $\{r_1,\cdots,r_n\}$ follows the uniform distribution on the permutations of integers $\{1,\cdots,n\}$. Consequently, $\vxi$ has a completely known distribution that is independent of the random error distribution.
Hence, the gradient function has the complete pivotal property  \citep{Parzen1994}, which implies the tuning-free
property of rank-Lasso. 
To see this, recall that
the general principal of tuning parameter selection 
\citep{bickel2009}
suggests choosing $\lambda$ such that
$P(\lambda>c||\vS_n||_\infty)\geq 1-\alpha_0$
for some constant $c > 1$ and a given small $\alpha_0>0$. With the design matrix $\vX$ and a completely known distribution of $\vxi$, we can easily simulate the distribution of $\vS_n$ and hence compute the theoretically optimal $\lambda$. 

\citet{Wang2018} established 
a finite-sample estimation error bound for the Rank Lasso estimator with the aforementioned simulated tuning parameter 
and showed that it achieves the same optimal near-oracle estimation error rate
as Lasso does. In contrast to Lasso, the conditions required by rank Lasso for the 
error distribution are much weaker and allow for heavy-tailed distributions such as Cauchy distribution. 
Moreover, they proved that further improvement in efficiency can be achieved
by a second-stage enhancement with some light tuning.

\section{Other alternative tuning parameter selection methods for Lasso}
\subsection{Bootstrap-based approach} 
\citet{Hall2009} developed an $m$-out-of-$n$ bootstrap algorithm to select the tuning parameter
for Lasso, pointing out that standard  bootstrap methods would fail. 
Their algorithm employs a wild bootstrap procedure (see Algorithm~\ref{algo:boots}), 
which allows one to estimate the mean squared error of the parameter estimators for different tuning parameters. 
For each candidate $\lambda$,
this algorithm computes the bootstrapped mean-square error estimate $\mbox{Err}(\lambda)$.
The optimal tuning parameter is chosen as
$
\widehat{\lambda}_{\scalebox{.7}{\mbox{boots}}} = \big(n/m\big)^{1/2}\argmin\limits_{\lambda} \mbox{Err}(\lambda).
$
The final estimator for $\vbeta_0$ is given by 
\bqa
\vbetah_{\scalebox{.7}{\mbox{boots}}} = \argmin\limits_{\vbeta}\Big\{  \sum_{i=1}^n (y_i-\bar{y}-\vx_i^T\vbeta)^2 + \widehat{\lambda}_{\scalebox{.7}{\mbox{boots}}}||\vbeta||_1\Big\}.
\eqa

 \begin{algorithm}[!h] 
 	\caption{Bootstrap algorithm} \label{algo:boots} 
 	\begin{algorithmic}[1]
 		\State Input $(\vy,\vX)$, a $\sqrt{n}$-consistent ``pilot estimator'' $\vbetaw$, and $\lambda$.
 		\State $\widehat{\epsilon}_i\leftarrow y_i-\bar{y}-\vx_i^T\vbetaw$.
 		\State $\widetilde{\epsilon}_i\leftarrow \widehat{\epsilon}_i-n^{-1}\sum_{j=1}^n \widehat{\epsilon}_j$.
 		\State Set $ \mbox{Err}(\vlam) \leftarrow 0$.
 		\For {k = 1,...,N} 
 		\State Obtain $\epsilon_1^*,...,\epsilon_m^*$ by sampling randomly from $\widetilde{\epsilon}_1,...,\widetilde{\epsilon}_n$ with replacement.
 		\State $y_i^* \leftarrow \bar{y} + \vx_i^T\vbetaw + \epsilon_i^*$, $i=1,...,m$.
 		\State $\vbetah^*(\lambda) \leftarrow \argmin\limits_{\vbeta} \big\{\sum_{i=1}^m(y_i^*-\bar{y}^*-\vx_i^T\vbeta)^2 + \lambda||\vbeta||_1\big\}$.
 		\State $\mbox{Err}(\lambda) \leftarrow \mbox{Err}(\lambda) + ||\vbetah^*(\lambda)-\vbetaw||^2$.
 		\EndFor
 		\Return  $\mbox{Err}(\lambda) $.	
 	\end{algorithmic}
 \end{algorithm}

Their method and theory were mostly developed for the $p<n$ case. 
The algorithm requires that the covariates are centered at their empirical means
and that a $\sqrt{n}$-consistent ``pilot estimator'' $\vbetaw$ is available. 
\citet{Hall2009} proved that if $m = O(n/(\log n)^{1+\eta})$ for some $\eta>0$, then 
the estimator $\vbetah_{\scalebox{.7}{\mbox{boots}}}$ can identify the true model
with probability approaching one as $n\ra\infty$. 
%in addition, this estimator is asymptotically normal with convergence rate $n^{-1/2}$. 
They also suggested that the theory can be generalized to the high dimensional case with fixed sparsity $||\vbeta_0||_0$, however, the order of $p$ would depend on the “generalized parameters” of the model such as the tail behaviors of the random noise.

\citet{chatterjee2011} proposed a modified bootstrap method for Lasso. 
This method first computes a thresholded version
of the Lasso estimator and then applies the residual bootstrap.
In the classical $p\ll n$ setting, \citet{chatterjee2011} proved that the modified bootstrap method  
provides valid approximation to the distribution of the
Lasso estimator. They further recommended to choose $\lambda$
to minimize the bootstrapped approximation to the mean squared error of the Lasso estimator.

\subsection{Adaptive calibration for $l_{\infty}$} 
Motivated by Lepski's method for non-parametric regression \citep{lepski1990,lepski1997},
\citet{Chichignoud2016} proposed a novel adaptive validation method
for tuning parameter selection for Lasso.
The method, named Adaptive Calibration for $l_{\infty}$ (AV$_\infty$),
performs simple tests along a single Lasso path to select the optimal tuning parameter.
The method is equipped with a fast computational routine and theoretical guarantees on
its finite-sample performance with
respect to the super-norm loss.

Let $\Lambda = \{\lambda_1,...,\lambda_N\}$ be a set of candidate values for $\lambda$, where $0<\lambda_1<\cdots<\lambda_N=\lambda_{\max} = 2n^{-1}||\vX^T\vy||_\infty$. 
Denote $\vbetah^{\scalebox{.7}{\mbox{Lasso}}}(\lambda_j)$ as the Lasso estimator in (\ref{Lasso}) with tuning parameter set as $\lambda=\lambda_j$, $j=1,\cdots,N$. The proposed AV$_\infty$
selects $\lambda$ based on the tests for sup-norm differences of Lasso estimates with different tuning parameters. It is defined as
 \begin{align}
\widehat{\lambda}_{\scalebox{.7}{\mbox{AC}}} = \min\Big\{\lambda\in\Lambda: \max\limits_{\substack{\lambda',\lambda''\in \Lambda\\ \lambda',\lambda''\geq \lambda}}\Big[ \frac{||\vbetah^{\scalebox{.7}{\mbox{Lasso}}}(\lambda')- \vbetah^{\scalebox{.7}{\mbox{Lasso}}}(\lambda'')||_\infty}{\lambda'+\lambda''}-\bar{C}\Big]\leq 0 \Big\},\label{AV_crit}
\end{align}
where $\bar{C}$ is a constant with respect to the $L_\infty$ error bound of Lasso estimator. \citet{Chichignoud2016} recommended the universal choice $\bar{C} = 0.75$ for all practical purposes. 

\citet{Chichignoud2016} proposed a simple and fast implementation for the tuning parameter selection via AV$_\infty$,
see the description in
Algorithm~\ref{algo:AV}, where in the algorithm the binary random variable $\widehat{t}_{\lambda_j}$ is defined as
\bqa
\widehat{t}_{\lambda_j} = \prod_{k=j}^{N}\mathbbm{1}\Big\{ \frac{||\vbetah(\lambda_j)- \vbetah(\lambda_k)||_\infty}{\lambda_j+\lambda_k}\leq\bar{C}\Big\},
\quad j=1, \ldots, N,
\eqa
with $\mathbbm{1}$ being the indicator function.
The final estimator for the AV$_\infty$ method is the Lasso estimator with the tuning parameter $\widehat{\lambda}_{\scalebox{.7}{\mbox{AC}}}$, denoted as $\vbetah(\widehat{\lambda}_{\scalebox{.7}{\mbox{AC}}})$. 
As shown in Algorithm~\ref{algo:AV}, AV$_\infty$ only needs to compute one solution path, in contrast to the $K$ paths in 
the $K$-fold cross-validation for Lasso. 
The new method is usually faster than cross-validation.
\citet{Chichignoud2016} proved that  $||\vbetah(\widehat{\lambda}_{\scalebox{.7}{\mbox{AC}}})-\vbeta_0||_\infty$ achieves the optimal
sup-norm error bound of Lasso up to a constant pre-factor with high probability under some regularity conditions.

\begin{algorithm}[!h] 
	\caption{AV$_\infty$ algorithm} \label{algo:AV} 
	\begin{algorithmic}[1]
		\State Input $\vbetah(\lambda_1),...,\vbetah(\lambda_N),\bar{C}$.
		\State Set $j\leftarrow N$.
		\While {$\widehat{t}_{\lambda_{j-1}}\neq 0$ and $j>1$} 
		\State  Update index $j\leftarrow j-1$.	
		\EndWhile
		\Return  $\widehat{\lambda} \leftarrow \lambda_j$.	
	\end{algorithmic}
\end{algorithm}

%under the assumption that $||\vbetah(\lambda)-\vbeta_0||_\infty\leq C\lambda$ for some constant $C$ when conditioned on the event %$$\mathcal{T}_\lambda = \Big\{\frac{\sigma||\vX^T\vepsilon||_\infty}{n}\leq \frac{\lambda}{4}\Big\}.$$ 
%This assumption can be converted to the $L_\infty$-restricted eigenvalue condition on the design matrix $\vX$.

\section{Nonconvex penalized high-dimensional regression and tuning for support recovery}
\subsection{Background}
Lasso is known to achieve accurate 
prediction under rather weak conditions \citep{Greenshtein2004}. However,
it is also widely recognized that Lasso 
requires stringent conditions on the design matrix $\vX$ to 
achieve variable selection consistency \citep{Zou2006, Zhao2006}.
In many scientific problems, it is of importance to identify relevant or active variables.
For example, biologists are often interested in identifying the genes associated with certain disease.
This problem is often referred to as {\it support recovery}, with the goal to identify
$\mathcal{S}_0=\{j: \beta_{0j}\neq 0, j=1, \ldots, p\}$.

\begin{figure}
	\centering
	\includegraphics[scale=0.4]{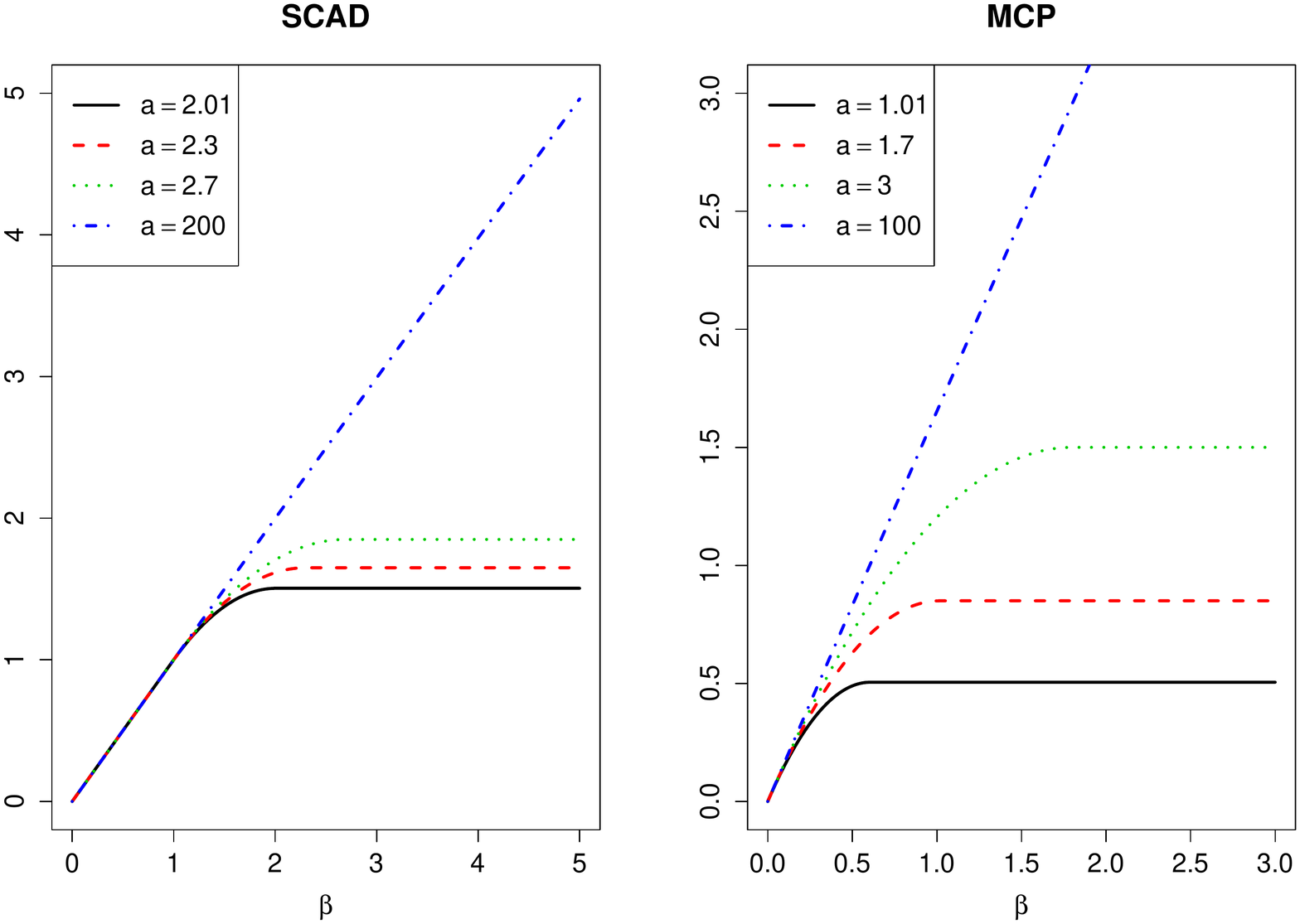}
	\caption{SCAD and MCP penalty functions ($\lambda=1$)}
	\label{fig:nonconvex}
\end{figure}

To alleviate the bias of Lasso due to 
the over-penalization of $L_1$ penalty, nonconvex penalized regression
has been studied in the literature as an alternative to Lasso \citep{FLreview, zhang2012review}. 
Two popular choices of nonconvex penalty functions
are SCAD \citep{FL:2001} and MCP \citep{Zhang:2010}. 
The SCAD penalty function is given by
\begin{align}
\label{scad}
p_{\lambda}(|\beta_j|) = \begin{cases}
\lambda |\beta_j|, &\mbox{ if } |\beta_j|\leq \lambda,\\
\frac{2a \lambda |\beta_j| -  \beta_j^2-\lambda^2}{2(a-1)},&\mbox{ if } \lambda<|\beta_j|< a\lambda,\\
\frac{(a+1)\lambda^2}{2},&\mbox{ if } |\beta_j|\geq a\lambda,
\end{cases}
\end{align}
where $a>2$ is a constant and \citet{FL:2001} recommended the choice $a = 3.7$.
The MCP penalty function is given by
\begin{align}
\label{mcp}
p_{\lambda}(|\beta_j|) = \begin{cases}
\lambda |\beta_j|-\frac{\beta_j^2}{2a}, &\mbox{ if } |\beta_j|\leq a\lambda,\\
\frac{a\lambda^2}{2},&\mbox{ if } |\beta_j|> a\lambda.
\end{cases}
\end{align}
where $a>1$ is a constant.
Figure~\ref{fig:nonconvex} depicts the two penalty functions.

As cross-validation for Lasso aims for prediction accuracy, it tends
to select a somewhat smaller tuning parameter (i.e., less regulation).
The resulted model size hence is usually larger than the true model size.
In the fixed $p$ setting, \citet{Wang2007} proved that with a positive probability cross-validation leads to a 
tuning parameter that would yield an over-fitted model.

Recent research has shown that when nonconvex penalized regression is combined with some modified BIC-type
criterion, the underlying model can be identified with probability approaching one under appropriate regularity conditions.
Several useful results were obtained in the low-dimensional setting.
For example, effective Bayesian information criterion (BIC) type
criterion for tuning parameter selection for nonconvex penalized regression
was investigated in Wang, Li and Tsai (2007) for fixed $p$ 
and Wang, Li and Leng (2009) for diverging $p$ (but $p<n$). 
\citet{zou2007degrees} considered Akaike information criterion (AIC) and BIC type criterion based on the degrees
of freedom for Lasso.
Also in the fixed $p$ setting, \citet{Zhang2010} studied generalized
information criterion, encompassing AIC and BIC.
They revealed that BIC-type selector enables identification of
the true model consistently and that AIC-type selector is asymptotically loss
efficient.

In the rest of this section, we review several modified BIC-type criteria in the high-dimensional setup ($p\gg n$) 
for tuning parameter selection with the goal of support recovery.

\subsection{Extended BIC for comparing models when $p\gg n$}
Let $\mathcal{S}$ be an arbitrary subset of $\{1,\cdots,p\}$. Hence, each $\mathcal{S}$
indexes a candidate model.  Given the data ($\vX,\vy$), 
the classical BIC, proposed by \citet{schwarz1978},  is defined as follows
$$\mbox{BIC}(\mathcal{S}) = -2\log L_n\{\vbetah(\mathcal{S})\} + ||\mathcal{S}||_0\log n,$$
where $L_n(\cdot)$ is the likelihood function,
$\vbetah(\mathcal{S})$ is the maximum likelihood estimator for the model with support $\mathcal{S}$,
and $||\mathcal{S}||_0$ is the cardinality of the set $\mathcal{S}$.
Given different candidate models, BIC selects the model with support $\mathcal{S}$ such that
BIC($\mathcal{S}$) is minimized. 

In the classical framework where $p$ is small and fixed, it is known \citep{Rao1989} that under standard conditions 
BIC is variable selection consistent, i.e., $\mathcal{S}_0$ is identified with probability approaching one 
as $n\ra \infty$ if the true model is in the set of candidate models. 
However, in the large $p$ setting, the number of candidate models grows exponentially fast in $p$.
The classical BIC is no longer computationally feasible.

\citet{chen2008} was the first to rigorously study the extension of BIC to high-dimensional regression where $p\gg n$.
They proposed an extended family of BIC of the form 
\begin{align}
\mbox{BIC}_\gamma(\mathcal{S}) = -2\log L_n\{\vbetah(\mathcal{S})\} + ||\mathcal{S}||_0\log n + 2\gamma\log \binom p{||\mathcal{S}||_0},\label{BICchen}
\end{align}
where $\gamma\in[0,1]$. Comparing with the classical BIC, the above modification incorporates the model size in the
penalty term.
It was proved that if $p=O(n^\kappa)$ for some constant $\kappa$, and $\gamma > 1 -(2\kappa)^{-1}$, then this extended BIC is
variable selection consistent under some regularity conditions.
\citet{kim2012consistent} also investigated variants of extended BIC
for comparing models for high-dimensional least-squares regression.

\subsection{HBIC for tuning parameter selection and support recovery}
The extended BIC is most useful if a candidate set of models is provided and if the true model is contained in such a candidate set
with high probability. One practical choice is to construct such a set of candidate models
from a Lasso solution path. As Lasso requires stringent conditions on the design matrix $\vX$
to be variable selection consistent. It is usually not guaranteed that the Lasso solution path contains the oracle
estimator, the estimator corresponding to support set $\mathcal{S}_0$.
Alternatively, one may construct a set of candidate models from the solution path of SCAD or MCP. 
However, as the objective function of SCAD or MCP is nonconvex, 
multiple minima may be present. The solution path of SCAD or MCP hence
may be nonunique and do not necessarily contain the oracle estimator. 
Even if a solution path is known to contain the oracle estimator, 
to find the optimal tuning parameter which yields the oracle estimator
with theoretical guarantee is challenging in high dimension.

To overcome these difficulties, 
\citet{wang2013} thoroughly studied how to calibrate non-convex penalized least squares 
regression to find the optimal tuning parameter for support recovery when $p\gg n$.
Define a consistent solution path to be a path that contains the oracle estimator with probability approaching one.
\citet{wang2013} first proved that an easy-to-calculate calibrated CCCP (CCCP stands for ConCave Convex procedure) 
algorithm produces a consistent solution path.
Furthermore, they proposed HBIC, a high-dimensional BIC criterion, and proved that it can be applied
to the solution path to select the optimal tuning parameter which
asymptotically identifies the oracle estimator. 
%Specifically, the tuning parameter $\lambda$ is selected by minimizing
%\begin{align}
%\mbox{BIC}(\lambda) = \log \{\frac{1}{n}||\vy-\vX \vbetah(\lambda)||^2\}+ ||\vbetah(\lambda)||_0 \frac{\log n}{n},\label{BICwang}
%\end{align}
Let $\widetilde{\vbeta}(\lambda)$ be
the solution corresponding to $\lambda$ on a consistent solution path, for example, the one obtained
by the aforementioned calibrated nonconvex-penalized regression with SCAD or MCP penalty. 
HBIC selects the optimal tuning parameter $\lambda$ in  $\Lambda_n=\{\lambda: ||\widetilde{\vbeta}(\lambda)||_0\leq K_n \}$, where $K_n$ is allowed to diverge to infinity, by minimizing
\begin{align}
\mbox{HBIC}(\lambda) = \log \Big\{\frac{1}{n}||\vy-\vX \widetilde{\vbeta}(\lambda)||^2\Big\} + ||\widetilde{\vbeta}(\lambda)||_0 \frac{C_n\log p}{n},\label{HBIC}
\end{align}
where $C_n$ diverges to infinity.
\citet{wang2013} proves that if $C_n ||\vbeta_0||_0\log p= o(n)$ and $K_n^2\log p\log n =o(n)$, then under mild conditions,  HBIC  identifies the true model with probability approaching one. For example, one can take $C_n=\log(\log n)$. Note that
the consistency is valid in the ultra-high dimensional setting, where $p$ is allowed to grow exponentially fast in $n$.

In addition, \citet{wang2011consistent} studied a variant of HBIC in combination of a sure screening procedure.
\citet{Fan2013} investigated 
proxy generalized information criterion,
a proxy of the generalized information criterion \citep{Zhang2010} when $p\gg n$.  
They identified a range
of complexity penalty levels such that the tuning parameter that is selected by optimizing the proxy generalized information
criterion can achieve model selection consistency.

\section{A real data example}
We consider the data set  sp500 in the R package \texttt{scalreg}, which contains a year's worth of close-of-day data for most of the Standard and Poors 500 stocks. 
The response variable \texttt{sp500.percent} is the daily percentage change.
The data set has 252 observations of 492 variables.

We demonstrate the performance of Lasso with $K$-fold cross validation, scaled Lasso and $\sqrt{\mbox{Lasso}}$ methods
on this example. Other methods reviewed in this paper which do not yet have publicly available software packages 
are not implemented. We evaluate the performance of different methods based on 100 random splits.
For each split, we randomly select half of the data to train the model, and 
then compute the $L_1$ and $L_2$-prediction errors and estimated model sizes
on the other half of the data. For Lasso, we select the tuning parameter by $10$-fold cross validation, using the 
R function ``cv.glmnet'' and the one-standard-error rule. For scaled Lasso, we apply the default tuning parameter selection
method in the R function ``scalreg'', which is  the quantile-based penalty level (\bfblue{lam0=``quantile"}) introduced and studied in \citet{Sun2013}. For $\sqrt{\mbox{Lasso}}$ method, we use R function ``slim'' to train the model. However, the package does not have a build-in tuning 
parameter selection method. As the optimal tuning parameter depends on the tail behavior of the random error,
it is also chosen by 10-fold cross validation.
 
Table~\ref{realdata} summarizes the averages and standard deviations of the $L_1$ and $L_2$- prediction errors 
and estimated model sizes for the three methods with 100 random splits. Lasso and $\sqrt{\mbox{Lasso}}$ have similar performance, though Lasso method tends to yield sparser models. Scaled Lasso has slightly larger prediction errors and model sizes.
The difference may be due to the non-normality of th data, which would affect the performance of the default tuning parameter selection 
method in the ``scalreg'' function.
  
\begin{table}[!h]
\centering
\caption{Analysis of sp500 data}\label{realdata}
\begin{tabular}{c|ccc}
	\hline 
	& Lasso & Scaled Lasso & $\sqrt{\mbox{Lasso}}$ \\  	\hline 
	$L_1$ error & 0.17 (0.02) & 0.21 (0.02) & 0.17 (0.02) \\  
	$L_2$ error & 0.05 (0.01) & 0.08 (0.03) & 0.05 (0.01) \\  
	Sparsity & 60.03 (5.39) & 120.82 (4.70) & 76.63 (8.27) \\ 
	\hline 
\end{tabular} 
\end{table}

\section{Discussions}
Developing computationally efficient tuning parameter selection methods with theoretical guarantees
is important for many high-dimensional statistical problems but has so far only received limited attention
in the current literature.
This paper reviews several commonly used tuning parameter selection approaches
for high-dimensional linear regression and provides some insights on how they work. 
The aim is to bring more attention to this important
topic to help stimulate future fruitful research in this direction.

The review article focused on regularized least squares types of estimation procedures
for sparse linear regression.
The specific choice of tuning parameter necessarily depends on the user's own research objectives:
Is prediction the main research goal? Or is identifying relevant variables of more importance?
How much computational time is the researcher willing to allocate? Is robustness of any concern
for the data set under consideration?  

The problem of tuning parameter selection is ubiquitous and has been investigated in
settings beyond sparse linear least squares regression.
\citet{Lee2014} extended the idea of extended BIC to high-dimensional quantile regression. They recommended to select the model that minimizes
\begin{align}
\mbox{BIC}_{\mbox{Q}}(\mathcal{S}) = \log\Big\{\sum_{i=1}^n\rho_\tau\big(y_i-\vx_{i}^T \vbetah(\mathcal{S})\big) \Big\} + (2n)^{-1}C_n ||\mathcal{S}||_0\log n,\label{QHBIC}
\end{align}
where $\rho_\tau(u) = 2u\big(\tau-I(u<0)\big)$ is the quantile loss function, and $C_n$ is some positive constant that diverges to infinity as $n$ increases. They also proved variable selection consistency property when $C_n\log n / n\ra 0$ under some regularity conditions. 
\citet{BC2011} and  \citet{koenker2011} considered tuning parameter selection for penalized quantile regression based
on the pivotal property of the quantile score function.\citet{wang2012} considered tuning parameter selection using 
cross-validation with the quantile loss function. For support vector machines (SVM), a widely used approach for classification,
\citet{zhang2016consistent} recently established the consistency of extended BIC type criterion for tuning
parameter selection in the high-dimensional setting.
For semiparametric regression models, \citet{xie2009scad} explored cross-validation for high-dimensional partially linear mean regression;
\citet{sherwood2016partially} applied an extended BIC type criterion for high-dimensional partially linear additive quantile regression.
\citet{datta2017cocolasso} derived a corrected cross-validation procedure for high-dimensional
linear regression with error in variables. 
In \citet{guo2016high}, an extended BIC type criterion was used for 
high-dimensional and banded vector autoregressions. In studying high-dimensional panel data,
\citet{kock2013oracle} empirically investigated both cross validation and BIC for tuning parameter selection.

Although the basic ideas of cross validation and BIC can be intuitively generalized to more complex
modeling settings, their theoretical justifications are often still lacking despite the promising numerical
evidence. It is worth emphasizing that intuition is not always straightforward 
and theoretical insights can be valuable. For instance, when investigating 
high-dimensional graphs and variable selection with the lasso,
\citet{meinshausen2006high} observed that
the consistency of neighborhood selection hinges on the choice of the penalty parameter. 
The oracle value for optimal prediction does not lead to a consistent
neighborhood estimate. 

\bibliography{Reviewref_071419}

\begin{thebibliography}{76}
\providecommand{\natexlab}[1]{#1}
\providecommand{\url}[1]{\texttt{#1}}
\expandafter\ifx\csname urlstyle\endcsname\relax
  \providecommand{\doi}[1]{doi: #1}\else
  \providecommand{\doi}{doi: \begingroup \urlstyle{rm}\Url}\fi

\bibitem[Antoniadis(2010)]{antoniadis2010comments}
A.~Antoniadis.
\newblock Comments on: {$l_1 $}-penalization for mixture regression models.
\newblock \emph{Test}, 19\penalty0 (2):\penalty0 257--258, 2010.

\bibitem[Belloni and Chernozhukov(2011)]{BC2011}
A.~Belloni and V.~Chernozhukov.
\newblock L1-penalized quantile regression in high-dimensional sparse models.
\newblock \emph{The Annals of Statistics}, 39:\penalty0 82--130, 2011.

\bibitem[Belloni et~al.(2011)Belloni, Chernozhukov, and Wang]{belloni2011}
A.~Belloni, V.~Chernozhukov, and L.~Wang.
\newblock Square-root lasso: pivotal recovery of sparse signals via conic
  programming.
\newblock \emph{Biometrika}, 98\penalty0 (4):\penalty0 791--806, 2011.

\bibitem[Bickel et~al.(2009)Bickel, Ritov, and Tsybakov]{bickel2009}
P.~J. Bickel, Y.~Ritov, and A.~B. Tsybakov.
\newblock Simultaneous analysis of lasso and dantzig selector.
\newblock \emph{The Annals of Statistics}, 37\penalty0 (4):\penalty0
  1705--1732, 2009.

\bibitem[Bien et~al.(2016)Bien, Gaynanova, Lederer, and
  M{\"u}ller]{bien2016non}
J.~Bien, I.~Gaynanova, J.~Lederer, and C.~M{\"u}ller.
\newblock Non-convex global minimization and false discovery rate control for
  the trex.
\newblock \emph{arXiv preprint arXiv:1604.06815}, 2016.

\bibitem[Bien et~al.(2018)Bien, Gaynanova, Lederer, and
  M{\"u}ller]{bien2018prediction}
J.~Bien, I.~Gaynanova, J.~Lederer, and C.~L. M{\"u}ller.
\newblock Prediction error bounds for linear regression with the trex.
\newblock \emph{TEST}, pages 1--24, 2018.

\bibitem[Boyd and Vandenberghe(2004)]{boyd2004}
S.~Boyd and L.~Vandenberghe.
\newblock \emph{Convex Optimization}.
\newblock Cambridge University Press, New York, NY, USA, 2004.

\bibitem[B{\"u}hlmann and van~de Geer(2011)]{buhlmann2011}
P.~B{\"u}hlmann and S.~van~de Geer.
\newblock \emph{Statistics for high-dimensional data: methods, theory and
  applications}.
\newblock Springer Science \& Business Media, 2011.

\bibitem[Bunea et~al.(2007)Bunea, Tsybakov, Wegkamp, et~al.]{bunea2007}
F.~Bunea, A.~Tsybakov, M.~Wegkamp, et~al.
\newblock Sparsity oracle inequalities for the lasso.
\newblock \emph{Electronic Journal of Statistics}, 1:\penalty0 169--194, 2007.

\bibitem[Candes and Tao(2007)]{candes2007dantzig}
E.~Candes and T.~Tao.
\newblock The dantzig selector: statistical estimation when p is much larger
  than n.
\newblock \emph{The Annals of Statistics}, 35\penalty0 (6):\penalty0
  2313--2351, 2007.

\bibitem[Cand{\`e}s et~al.(2009)Cand{\`e}s, Plan, et~al.]{Candes2009}
E.~J. Cand{\`e}s, Y.~Plan, et~al.
\newblock Near-ideal model selection by l1 minimization.
\newblock \emph{The Annals of Statistics}, 37\penalty0 (5A):\penalty0
  2145--2177, 2009.

\bibitem[Chatterjee and Lahiri(2011)]{chatterjee2011}
A.~Chatterjee and S.~N. Lahiri.
\newblock Bootstrapping lasso estimators.
\newblock \emph{Journal of the American Statistical Association}, 106\penalty0
  (494):\penalty0 608--625, 2011.

\bibitem[Chatterjee and Jafarov(2015)]{Chatterjee2015}
S.~Chatterjee and J.~Jafarov.
\newblock Prediction error of cross-validated lasso.
\newblock \emph{arXiv preprint arXiv:1502.06291}, 2015.

\bibitem[Chen and Chen(2008)]{chen2008}
J.~Chen and Z.~Chen.
\newblock {Extended Bayesian information criteria for model selection with
  large model spaces}.
\newblock \emph{Biometrika}, 95\penalty0 (3):\penalty0 759--771, 2008.

\bibitem[Chen et~al.(2001)Chen, Donoho, and Saunders]{chen2001}
S.~S. Chen, D.~L. Donoho, and M.~A. Saunders.
\newblock Atomic decomposition by basis pursuit.
\newblock \emph{SIAM review}, 43\penalty0 (1):\penalty0 129--159, 2001.

\bibitem[Chetverikov et~al.(2016)Chetverikov, Liao, and
  Chernozhukov]{Chetverikov2016}
D.~Chetverikov, Z.~Liao, and V.~Chernozhukov.
\newblock On cross-validated lasso.
\newblock \emph{arXiv preprint arXiv:1605.02214}, 2016.

\bibitem[Chichignoud et~al.(2016)Chichignoud, Lederer, and
  Wainwright]{Chichignoud2016}
M.~Chichignoud, J.~Lederer, and M.~Wainwright.
\newblock A practical scheme and fast algorithm to tune the lasso with
  optimality guarantees.
\newblock \emph{Journal of Machine Learning Research}, 17:\penalty0 1--20, 12
  2016.

\bibitem[Datta et~al.(2017)Datta, Zou, et~al.]{datta2017cocolasso}
A.~Datta, H.~Zou, et~al.
\newblock Cocolasso for high-dimensional error-in-variables regression.
\newblock \emph{The Annals of Statistics}, 45\penalty0 (6):\penalty0
  2400--2426, 2017.

\bibitem[Efron et~al.(2004)Efron, Hastie, Johnstone, and
  Tibshirani]{efron2004least}
B.~Efron, T.~Hastie, I.~Johnstone, and R.~Tibshirani.
\newblock Least angle regression.
\newblock \emph{The Annals of statistics}, 32\penalty0 (2):\penalty0 407--499,
  2004.

\bibitem[Fan and Li(2001)]{FL:2001}
J.~Fan and R.~Li.
\newblock Variable selection via nonconcave penalized likelihood and its oracle
  property.
\newblock \emph{Journal of the American Statistical Association}, 96:\penalty0
  1348--1360, 2001.

\bibitem[Fan and Lv(2010)]{FLreview}
J.~Fan and J.~Lv.
\newblock A selective overview of variable selection in high dimensional
  feature space.
\newblock \emph{Statistica Sinica}, 20\penalty0 (1):\penalty0 101, 2010.

\bibitem[Fan and Tang(2013)]{Fan2013}
Y.~Fan and C.~Y. Tang.
\newblock Tuning parameter selection in high dimensional penalized likelihood.
\newblock \emph{Journal of the Royal Statistical Society: Series B},
  75\penalty0 (3):\penalty0 531--552, 2013.

\bibitem[Friedman et~al.(2007)Friedman, Hastie, H{\"o}fling, and
  Tibshirani]{friedman2007pathwise}
J.~Friedman, T.~Hastie, H.~H{\"o}fling, and R.~Tibshirani.
\newblock Pathwise coordinate optimization.
\newblock \emph{The Annals of Applied Statistics}, 1\penalty0 (2):\penalty0
  302--332, 2007.

\bibitem[Friedman et~al.(2010)Friedman, Hastie, and Tibshirani]{glmnet}
J.~Friedman, T.~Hastie, and R.~Tibshirani.
\newblock Regularization paths for generalized linear models via coordinate
  descent.
\newblock \emph{Journal of Statistical Software}, 33\penalty0 (1):\penalty0
  1--22, 2010.

\bibitem[Greenshtein and Ritov(2004)]{Greenshtein2004}
E.~Greenshtein and Y.~Ritov.
\newblock Persistence in high-dimensional linear predictor selection and the
  virtue of overparametrization.
\newblock \emph{Bernoulli}, 10\penalty0 (6):\penalty0 971--988, 2004.

\bibitem[Guo et~al.(2016)Guo, Wang, and Yao]{guo2016high}
S.~Guo, Y.~Wang, and Q.~Yao.
\newblock High-dimensional and banded vector autoregressions.
\newblock \emph{Biometrika}, page asw046, 2016.

\bibitem[Hall et~al.(2009)Hall, Lee, and Park]{Hall2009}
P.~Hall, E.~R. Lee, and B.~U. Park.
\newblock Bootstrap-based penalty choice for the lasso, achieving oracle
  performance.
\newblock \emph{Statistica Sinica}, 19\penalty0 (2):\penalty0 449--471, 2009.

\bibitem[Hastie and Efron(2013)]{lars}
T.~Hastie and B.~Efron.
\newblock \emph{lars: Least Angle Regression, Lasso and Forward Stagewise},
  2013.
\newblock URL \url{https://CRAN.R-project.org/package=lars}.
\newblock R package version 1.2.

\bibitem[Hastie et~al.(2015)Hastie, Tibshirani, and
  Wainwright]{hastie2015statistical}
T.~Hastie, R.~Tibshirani, and M.~Wainwright.
\newblock \emph{Statistical learning with sparsity: the lasso and
  generalizations}.
\newblock Chapman and Hall/CRC, 2015.

\bibitem[Hastie and Friedman(2009)]{Hastie09}
T.~R. Hastie, T. and J.~Friedman.
\newblock \emph{The Elements of Statistical Learning: Data Mining, Inference,
  and Prediction (2nd ed.)}.
\newblock New York: Springer., 2009.

\bibitem[Homrighausen and McDonald(2013)]{Homrighausen2013}
D.~Homrighausen and D.~McDonald.
\newblock The lasso, persistence, and cross-validation.
\newblock In \emph{International Conference on Machine Learning}, pages
  1031--1039, 2013.

\bibitem[Homrighausen and McDonald(2017)]{Homrighausen2017}
D.~Homrighausen and D.~J. McDonald.
\newblock Risk consistency of cross-validation with lasso-type procedures.
\newblock \emph{Statistica Sinica}, pages 1017--1036, 2017.

\bibitem[Jaeckel(1972)]{jaeckel1972}
L.~A. Jaeckel.
\newblock Estimating regression coefficients by minimizing the dispersion of
  the residuals.
\newblock \emph{The Annals of Mathematical Statistics}, 43\penalty0
  (5):\penalty0 1449--1458, 1972.

\bibitem[Kim et~al.(2012)Kim, Kwon, and Choi]{kim2012consistent}
Y.~Kim, S.~Kwon, and H.~Choi.
\newblock Consistent model selection criteria on high dimensions.
\newblock \emph{Journal of Machine Learning Research}, 13\penalty0
  (Apr):\penalty0 1037--1057, 2012.

\bibitem[Kock(2013)]{kock2013oracle}
A.~B. Kock.
\newblock Oracle efficient variable selection in random and fixed effects panel
  data models.
\newblock \emph{Econometric Theory}, 29\penalty0 (1):\penalty0 115--152, 2013.

\bibitem[Koenker(2011)]{koenker2011}
R.~Koenker.
\newblock Additive models for quantile regression: Model selection and
  confidence bandaids.
\newblock \emph{Brazilian Journal of Probability and Statistics}, 25\penalty0
  (3):\penalty0 239--262, 2011.

\bibitem[Lederer and Müller(2015)]{lederer2015}
J.~Lederer and C.~Müller.
\newblock Don't fall for tuning parameters: Tuning-free variable selection in
  high dimensions with the trex, 2015.

\bibitem[Lee et~al.(2014)Lee, Noh, and Park]{Lee2014}
E.~R. Lee, H.~Noh, and B.~U. Park.
\newblock Model selection via bayesian information criterion for quantile
  regression models.
\newblock \emph{Journal of the American Statistical Association}, 109\penalty0
  (505):\penalty0 216--229, 2014.

\bibitem[Lepski(1991)]{lepski1990}
O.~Lepski.
\newblock On a problem of adaptive estimation in gaussian white noise.
\newblock \emph{Theory of Probability \& Its Applications}, 35\penalty0
  (3):\penalty0 454--466, 1991.

\bibitem[Lepski and Spokoiny(1997)]{lepski1997}
O.~V. Lepski and V.~G. Spokoiny.
\newblock Optimal pointwise adaptive methods in nonparametric estimation.
\newblock \emph{The Annals of Statistics}, 25\penalty0 (6):\penalty0
  2512--2546, 1997.

\bibitem[Li et~al.(2018)Li, Zhao, Wang, Yuan, and Liu]{flare}
X.~Li, T.~Zhao, L.~Wang, X.~Yuan, and H.~Liu.
\newblock \emph{flare: Family of Lasso Regression}, 2018.
\newblock URL \url{https://CRAN.R-project.org/package=flare}.
\newblock R package version 1.6.0.

\bibitem[Meinshausen and B{\"u}hlmann(2006)]{meinshausen2006high}
N.~Meinshausen and P.~B{\"u}hlmann.
\newblock High-dimensional graphs and variable selection with the lasso.
\newblock \emph{The annals of statistics}, pages 1436--1462, 2006.

\bibitem[Negahban et~al.(2012)Negahban, Ravikumar, Wainwright, Yu,
  et~al.]{Negahban2012}
S.~N. Negahban, P.~Ravikumar, M.~J. Wainwright, B.~Yu, et~al.
\newblock A unified framework for high-dimensional analysis of $m$-estimators
  with decomposable regularizers.
\newblock \emph{Statistical Science}, 27\penalty0 (4):\penalty0 538--557, 2012.

\bibitem[Owen(2007)]{Owen2007}
A.~B. Owen.
\newblock A robust hybrid of lasso and ridge regression.
\newblock 2007.

\bibitem[Parzen et~al.(1994)Parzen, Wei, and Ying]{Parzen1994}
M.~Parzen, L.~Wei, and Z.~Ying.
\newblock A resampling method based on pivotal estimating functions.
\newblock \emph{Biometrika}, 81\penalty0 (2):\penalty0 341--350, 1994.

\bibitem[Rao and Wu(1989)]{Rao1989}
R.~Rao and Y.~Wu.
\newblock A strongly consistent procedure for model selection in a regression
  problem.
\newblock \emph{Biometrika}, 76\penalty0 (2):\penalty0 369--374, 1989.

\bibitem[Schwarz(1978)]{schwarz1978}
G.~Schwarz.
\newblock Estimating the dimension of a model.
\newblock \emph{Ann. Statist.}, 6\penalty0 (2):\penalty0 461--464, 03 1978.

\bibitem[Sherwood et~al.(2016)Sherwood, Wang, et~al.]{sherwood2016partially}
B.~Sherwood, L.~Wang, et~al.
\newblock Partially linear additive quantile regression in ultra-high
  dimension.
\newblock \emph{The Annals of Statistics}, 44\penalty0 (1):\penalty0 288--317,
  2016.

\bibitem[St{\"a}dler et~al.(2010)St{\"a}dler, B{\"u}hlmann, and Van
  De~Geer]{stadler2010}
N.~St{\"a}dler, P.~B{\"u}hlmann, and S.~Van De~Geer.
\newblock {$l_1 $}-penalization for mixture regression models.
\newblock \emph{Test}, 19\penalty0 (2):\penalty0 209--256, 2010.

\bibitem[Sun and Zhang(2010)]{sun2010comments}
T.~Sun and C.-H. Zhang.
\newblock Comments on: {$l_1 $}-penalization for mixture regression models.
\newblock \emph{Test}, 19\penalty0 (2):\penalty0 270--275, 2010.

\bibitem[Sun and Zhang(2012)]{sun2012scaled}
T.~Sun and C.-H. Zhang.
\newblock Scaled sparse linear regression.
\newblock \emph{Biometrika}, 99\penalty0 (4):\penalty0 879--898, 2012.

\bibitem[Sun and Zhang(2013)]{Sun2013}
T.~Sun and C.-H. Zhang.
\newblock Sparse matrix inversion with scaled lasso.
\newblock \emph{J. Mach. Learn. Res.}, 14\penalty0 (1):\penalty0 3385--3418,
  Jan. 2013.
\newblock ISSN 1532-4435.
\newblock URL \url{http://dl.acm.org/citation.cfm?id=2567709.2567771}.

\bibitem[Tian et~al.(2018)Tian, Loftus, and Taylor]{tian2018selective}
X.~Tian, J.~R. Loftus, and J.~E. Taylor.
\newblock Selective inference with unknown variance via the square-root lasso.
\newblock \emph{Biometrika}, 105\penalty0 (4):\penalty0 755--768, 2018.

\bibitem[Tibshirani(1996)]{tibshirani1996}
R.~Tibshirani.
\newblock Regression shrinkage and selection via the lasso.
\newblock \emph{Journal of the Royal Statistical Society. Series B
  (Methodological)}, 58\penalty0 (1):\penalty0 267--288, 1996.

\bibitem[Van~de Geer(2016)]{van2016estimation}
S.~Van~de Geer.
\newblock Estimation and testing under sparsity.
\newblock 2016.

\bibitem[Van~de Geer et~al.(2008)]{van2008}
S.~A. Van~de Geer et~al.
\newblock High-dimensional generalized linear models and the lasso.
\newblock \emph{The Annals of Statistics}, 36\penalty0 (2):\penalty0 614--645,
  2008.

\bibitem[Wainwright(2019)]{Wainwright19}
M.~Wainwright.
\newblock \emph{High-Dimensional Statistics: A Non-Asymptotic Viewpoint}.
\newblock Cambridge Series in Statistical and Probabilistic Mathematics, 2019.

\bibitem[Wang et~al.(2007)Wang, Li, and Tsai]{Wang2007}
H.~Wang, R.~Li, and C.-L. Tsai.
\newblock {Tuning parameter selectors for the smoothly clipped absolute
  deviation method}.
\newblock \emph{Biometrika}, 94\penalty0 (3):\penalty0 553--568, 2007.

\bibitem[Wang(2009)]{wang2009}
L.~Wang.
\newblock Wilcoxon-type generalized bayesian information criterion.
\newblock \emph{Biometrika}, 96\penalty0 (1):\penalty0 163--173, 2009.

\bibitem[Wang and Li(2009)]{wang2009local}
L.~Wang and R.~Li.
\newblock Weighted wilcoxon-type smoothly clipped absolute deviation method.
\newblock \emph{Biometrics}, 65\penalty0 (2):\penalty0 564--571, 2009.

\bibitem[Wang et~al.(2012)Wang, Wu, and Li]{wang2012}
L.~Wang, Y.~Wu, and R.~Li.
\newblock Quantile regression for analyzing heterogeneity in ultra-high
  dimension.
\newblock \emph{Journal of the American Statistical Association}, 107\penalty0
  (497):\penalty0 214--222, 2012.

\bibitem[Wang et~al.(2013)Wang, Kim, and Li]{wang2013}
L.~Wang, Y.~Kim, and R.~Li.
\newblock Calibrating nonconvex penalized regression in ultra-high dimension.
\newblock \emph{Ann. Statist.}, 41:\penalty0 2505--2536, 2013.

\bibitem[Wang et~al.(2018)Wang, Peng, Bradic, Li, and Wu]{Wang2018}
L.~Wang, B.~Peng, J.~Bradic, R.~Li, and Y.~Wu.
\newblock A tuning-free robust and efficient approach to high-dimensional
  regression.
\newblock Technical report, School of Statistics, University of Minnesota,
  2018.

\bibitem[Wang and Zhu(2011)]{wang2011consistent}
T.~Wang and L.~Zhu.
\newblock Consistent tuning parameter selection in high dimensional sparse
  linear regression.
\newblock \emph{Journal of Multivariate Analysis}, 102\penalty0 (7):\penalty0
  1141--1151, 2011.

\bibitem[Wu et~al.(2008)Wu, Lange, et~al.]{wu2008coordinate}
T.~T. Wu, K.~Lange, et~al.
\newblock Coordinate descent algorithms for lasso penalized regression.
\newblock \emph{The Annals of Applied Statistics}, 2\penalty0 (1):\penalty0
  224--244, 2008.

\bibitem[Xie et~al.(2009)Xie, Huang, et~al.]{xie2009scad}
H.~Xie, J.~Huang, et~al.
\newblock Scad-penalized regression in high-dimensional partially linear
  models.
\newblock \emph{The Annals of Statistics}, 37\penalty0 (2):\penalty0 673--696,
  2009.

\bibitem[Zhang(2010{\natexlab{a}})]{Zhang:2010}
C.~H. Zhang.
\newblock Nearly unbiased variable selection under minimax concave penalty.
\newblock \emph{Annals of Statistics}, 38:\penalty0 894--942,
  2010{\natexlab{a}}.

\bibitem[Zhang and Zhang(2012)]{zhang2012review}
C.-H. Zhang and T.~Zhang.
\newblock A general theory of concave regularization for high-dimensional
  sparse estimation problems.
\newblock \emph{Statistical Science}, pages 576--593, 2012.

\bibitem[Zhang et~al.(2008)Zhang, Huang, et~al.]{zhang2008sparsity}
C.-H. Zhang, J.~Huang, et~al.
\newblock The sparsity and bias of the lasso selection in high-dimensional
  linear regression.
\newblock \emph{The Annals of Statistics}, 36\penalty0 (4):\penalty0
  1567--1594, 2008.

\bibitem[Zhang(2010{\natexlab{b}})]{zhang2010analysis}
T.~Zhang.
\newblock Analysis of multi-stage convex relaxation for sparse regularization.
\newblock \emph{Journal of Machine Learning Research}, 11\penalty0
  (Mar):\penalty0 1081--1107, 2010{\natexlab{b}}.

\bibitem[Zhang et~al.(2016)Zhang, Wu, Wang, and Li]{zhang2016consistent}
X.~Zhang, Y.~Wu, L.~Wang, and R.~Li.
\newblock A consistent information criterion for support vector machines in
  diverging model spaces.
\newblock \emph{The Journal of Machine Learning Research}, 17\penalty0
  (1):\penalty0 466--491, 2016.

\bibitem[Zhang et~al.(2010)Zhang, Li, and Tsai]{Zhang2010}
Y.~Zhang, R.~Li, and C.-L. Tsai.
\newblock Regularization parameter selections via generalized information
  criterion.
\newblock \emph{Journal of the American Statistical Association}, 105\penalty0
  (489):\penalty0 312--323, 2010.

\bibitem[Zhao and Yu(2006{\natexlab{a}})]{ZY2006}
P.~Zhao and B.~Yu.
\newblock On model selection consistency of lasso.
\newblock \emph{Journal of Machine Learning Research}, 7:\penalty0 2541--2563,
  2006{\natexlab{a}}.

\bibitem[Zhao and Yu(2006{\natexlab{b}})]{Zhao2006}
P.~Zhao and B.~Yu.
\newblock On model selection consistency of lasso.
\newblock \emph{J. Mach. Learn. Res.}, 7:\penalty0 2541--2563,
  2006{\natexlab{b}}.
\newblock ISSN 1532-4435.

\bibitem[Zou(2006)]{Zou2006}
H.~Zou.
\newblock The adaptive lasso and its oracle properties.
\newblock \emph{Journal of the American Statistical Association}, 101\penalty0
  (476):\penalty0 1418--1429, 2006.

\bibitem[Zou et~al.(2007)Zou, Hastie, and Tibshirani]{zou2007degrees}
H.~Zou, T.~Hastie, and R.~Tibshirani.
\newblock On the “degrees of freedom” of the lasso.
\newblock \emph{The Annals of Statistics}, 35\penalty0 (5):\penalty0
  2173--2192, 2007.

\end{thebibliography}

\end{document}